\documentclass{ws-procs9x6}

\begin{document}

\title{Coupled channel study of $K^+\Lambda$ photoproduction }

\author{B. Julia-Diaz, B. Saghai}

\address{D\'epartement d'Astrophysique, de Physique des Particules,
de Physique Nucl\'eaire et de
l'Instrumentation Associ\'ee, DSM, CEA/Saclay, 91191 Gif-sur-Yvette,
France\\
E-mail: bruno.julia-diaz@cea.fr, bijan.saghai@cea.fr\\}

\author{T.-S.H. Lee\footnote{\uppercase{U.S.} \uppercase{D}epartment of 
\uppercase{E}nergy, \uppercase{O}ffice of \uppercase{N}uclear \uppercase{P}hysics, 
               \uppercase{C}ontract No. \uppercase{W-31-109-ENG-38.}}}

\address{Physics Division, Argonne National Laboratory, Argonne,
               IL 60439, USA\\
E-mail: lee@phy.anl.gov}

\author{F. Tabakin\footnote{\uppercase{N}ational \uppercase{S}cience \uppercase{F}oundation,
grant No. 0244526 at the \uppercase{U}niversity of \uppercase{P}ittsburgh.}}

\address{Department of Physics and Astronomy,\\
University of Pittsburgh, PA 15260, USA\\
E-mail: tabakin@pitt.edu}

\maketitle

\abstracts{A coupled channel model with $\gamma N$, $KY$ and 
$\pi N$ channels has been used to analyze the recent data of 
$\gamma p \to K^+ \Lambda$. The non-resonant interactions within 
the subspace $KY \oplus \pi N$ are derived from effective 
Lagrangians using a unitary transformation method. The direct 
photoproduction reaction is obtained from a chiral constituent quark model 
with $SU(6)\otimes O(3)$ breaking. Missing baryon resonances issues
are briefly discussed.}

\section{Introduction}

Our knowledge of associated strangeness photoproduction processes has been
greatly improved in recent years thanks to measurements performed 
at several facilities, JLAB~\cite{JLab0,JLab1}, 
ELSA~\cite{ELSA,SAPHIR05} and Spring-8~\cite{LEPS}. 
This database should serve to shed light on the properties of  
known, poorly known, and missing resonances.

Our main aim in this contribution is to analyze the reaction 
$\gamma p \to K^+\Lambda$ making use of a coupled channel formalism. 
This work is an extension of Refs.~[\refcite{CC-01,CC-04,CCc-05}] with 
improvements in the derivation of the meson-baryon intermediate states
($\pi N \to \pi N$, $\pi N \to K Y$, and $K Y \to K Y$) and making 
use of the quark model of Refs.~[\refcite{CQM,S11}] for the direct and resonance 
photoproduction of $K^+\Lambda$. A more detailed description of this 
investigation will be reported elsewhere~\cite{CC-inprep}.

The main physics under scrutiny are the properties of resonances. This 
analysis should improve our knowledge of the properties of more or less 
well known resonances and 
also possible manifestations of missing resonances, predicted by QCD-inspired
approaches~\cite{Review}. In our study 
a $3^{rd}$ $S_{11}$, $3^{rd}$ $P_{13}$ and $3^{rd}$ $D_{13}$ resonances 
are considered and their relevance is examined. 

\section{Theoretical model}

A simple glimpse at the cross sections for meson photoproduction 
reveals that the pion photoproduction process is orders of magnitude 
larger than for strange photoproduction~\cite{bl}. Thus, it is obvious 
that part of the strange production flux will come from 
first producing a $\pi N$ intermediate state which subsequently decays 
into the $KY$ system. A suitable method to account for these processes, as well
for the final state interactions, 
is to consider a coupled channel formalism which includes the most relevant 
opened channels in the considered regime. 

Studying the effects of coupled channels is however an 
involved task due to the many intermediate and final state channels which 
are active. Thus, we need to have reasonable models for the following 
mechanisms: $\gamma N \to \pi N$, $\pi N \to \pi N$, $\pi N \to KY$ and
$KY \to KY$ in the considered total center-of-mass energy regime, 
$W \approx 1.6-2.7$ GeV. The main sources utilized in this work are 
Ref.~[\refcite{SatoLee}] for $\pi N$ and Ref.~[\refcite{CC-04}] with 
a number of improvements in the formulation for the $KY$ hadronic channels. 

The procedure employed~\cite{CC-04} to fit the parameters involved in the model has 
been to first fix the meson-baryon model parameters performing a $\chi^2$ fit 
to the available $\pi N \to K Y$ data. 
Then the photoproduction process
has been studied keeping the meson-baryon part unaltered. 

The direct $K\Lambda$ 
photoproduction process is handled using the quark model of Refs.~[\refcite{CQM,S11}]. 
That approach allows one to include all 3 and 4 star resonances and contains 
one adjustable parameter per resonance with masses below 2 GeV, due to
the $SU(6) \otimes O(3)$ symmetry breaking.

A detailed description of the formalism will be published elsewhere~\cite{CC-inprep}.

\section{Results}

\begin{figure}[t]
\vspace{20pt}
\includegraphics[width=90mm]{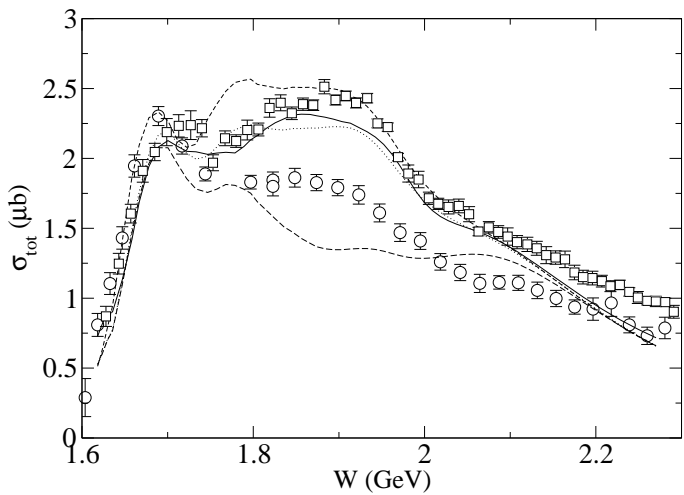}
\caption{Total cross section as a function of the 
centre of mass energy. Solid curve corresponds to the full model, dotted, 
dashed and long dashed correspond to switching off the $3^{rd}P_{13}$, 
$3^{rd}S_{11}$ and $3^{rd}D13$ resonances, respectively. 
Circle symbols, are SAPHIR data, 
Ref.~\protect\refcite{ELSA}. Square symbols correspond to the data 
from CLAS~\protect\cite{JLab0}.
\label{fig1}}
\end{figure}

The present experimental status is the following: there are two quite complete 
measurements of the differential cross section performed at CLAS~\cite{JLab0}
and SAPHIR~\cite{ELSA}. The 
most recent CLAS data~\cite{JLab0} for differential cross sections shows a 
closer agreement 
to SAPHIR data than those released~\cite{JLab1} previously. 
Secondly there 
are measured data for recoil polarization asymmetry from CLAS~\cite{JLab1} 
and also polarized photon asymmetry measured at LEPS~\cite{LEPS}.  

We have performed a thorough study of the compelete database~\cite{CC-inprep}. 
Fitting separately SAPHIR
and CLAS data, leads to reduced $\chi ^2$ of 1.3 and 2.1, respectively. 
Then, fitting {\it simultaneously} all cross section and polarization asymmetry data, 
we get $\chi_{d.o.f}^2 \approx 3$.

In fig.~\ref{fig1} the total cross section predicted by this model is compared  
to the data. To study the role played by new resonances introduced,
we also show results obtained by switching off those resonances one by one, without 
further fittings. Here, we emphasize that discrepencies between the two data sets are
much smaller in the differential cross sections than in the total cross sections
shown here. Larger discrepencies in the latter case are very likely due to
the angular range covered by each data set and extrapolation methods used to 
extract the total from the differential cross sections.

The figure summarizes our main findings which are to be presented in a longer 
discussion~\cite{CC-inprep}: the full model allows for a good reproduction 
of the data, the possible influence of the $3^{rd}P_{13}$ is very minor, and 
finally the role played by $3^{rd}S_{11}$ and $3^{rd}D_{13}$ are sizable. The mass 
and width of these resonances in this model are : $S_{11}$ [M=1.84 GeV, $\Gamma$=283 MeV] 
and $D_{13}$[M=1.93 GeV, $\Gamma$=252 MeV]. These findings are in line with other studied
with respect to the manifestations of a new $S_{11}$, Ref.~[\refcite{S11}], and a new 
$D_{13}$, Ref.~[\refcite{D13}], resonances.

\end{document}